\documentclass[prd,aps,twocolumn,nofootinbib,showpacs,preprintnumbers,amsmath,amssymb]{revtex4}
\usepackage[dvips]{graphicx}
\usepackage{amsmath}
\usepackage{amssymb}

\usepackage{graphicx}
\usepackage{dcolumn}
\usepackage{bm}
\def\be{\begin{equation}}
\def\ee{\end{equation}}
\def\bea{\begin{eqnarray}}
\def\eea{\end{eqnarray}}

\newcommand{\comment}[1]{}

\newcommand{\LF}{\left(}
\newcommand{\RF}{\right)}

\newcommand{\al}{\alpha}
\newcommand{\bb}{\beta}

\newcommand{\te}{\theta}

\newcommand{\de}{\delta}

\newcommand{\om}{\omega}

\newcommand{\n}{\nabla}

\newcommand{\caH}{{\cal H}}

\newcommand{\ra}{\rightarrow}
\newcommand{\Ra}{\Rightarrow}

\newcommand{\LT}{\left[}
\newcommand{\RT}{\right]}
\newcommand{\ie}{{\it i.e.\ }}

\newcommand{\cH}{{\cal H}}
\newcommand{\mx}{\mbox}

\begin{document}

\preprint{IGPG-07/7-1}

\date{\today}

\title{On the Transfer of Adiabatic Fluctuations through a
Nonsingular Cosmological Bounce}

\author{Stephon Alexander$^{1)}$ \email[email:sha3@psu.edu],
Tirthabir Biswas$^{1)}$ \email[email: ]{tbiswas@gravity.psu.edu}, and
Robert H. Brandenberger$^{2)}$ \email[email: ]{rhb@hep.physics.mcgill.ca}}

\affiliation{1) Department of Physics, Pennsylvania State University,
University Park, PA, 16802-6300, USA}

\affiliation{2) Department of Physics, McGill University,
Montr\'eal, QC, H3A 2T8, Canada}


\begin{abstract}

We study the transfer of cosmological perturbations through a
nonsingular cosmological bounce in a special model in which the
parameters of the bounce and the equation of state of matter are
chosen such as to allow for an exact calculation of the evolution of
the fluctuations. We find that the growing mode of the metric
fluctuations in the contracting phase goes over into the growing mode
in the expanding phase, a result which is different from what is
obtained in analyses in which fluctuations are matched at a singular
hypersurface. Consequences for Ekpyrotic cosmology are discussed in a
limit when the equation of state of a fluid becomes large.

\end{abstract}

\maketitle

\newcommand{\eq}[2]{\begin{equation}\label{#1}{#2}\end{equation}}

\section{Introduction}

It has long been appreciated that Standard Big Bang cosmology cannot
be the complete theory of the early universe since it is plagued by an
initial singularity \cite{Hawking}. The singularity theorems have also
been extended to apply to scalar field-driven inflationary cosmology
\cite{Borde}. It is possible that the initial singularity is resolved
by quantum gravity. Another possibility, however, is that correction
terms to the gravitational Lagrangian which describes the dynamics of
space-time leads to a bouncing universe. In a bouncing cosmology, the
Hubble radius $H^{-1}(t)$ (where $H$ is the Hubble expansion rate and
$t$ is physical time) decreases faster than the physical wavelength of
fluctuations (which have a constant wavelength in comoving
coordinates). Hence, it is conceivable that processes acting in the
contracting phase can lead to a non-inflationary mechanism for the
origin of structure in the universe. Recent proposals which involve
bouncing cosmologies are the Pre-Big-Bang \cite{PBB}, Ekpyrotic
\cite{EKP}, and  higher derivative modification of an Einstein
gravity~\cite{Biswas1} scenarios.

The transfer of metric fluctuations through the bounce from the
contracting to the expanding phase has been an outstanding problem in
the discussions of Pre-Big-Bang and Ekpyrotic cosmology. Often, the
evolution of the background is modeled by a contracting phase
modeled by a solution to the Einstein equations matched to an
expanding phase of Standard Big Bang cosmology through an
instantaneous and often singular transition along a space-like
hypersurface. One proposal has been to use the analog of the Israel
matching conditions (matching conditions \cite{Israel} which describe
the merger of two solutions of the Einstein equations along a
time-like hypersurface).  These equations were discussed in
\cite{Hwang,Deruelle}.

If we consider Einstein gravity, then in Fourier space the space of
solutions for fixed comoving wavenumber $k$ is two-dimensional (see
e.g. \cite{MFB} for a comprehensive review of the theory of
cosmological perturbations and \cite{RHBrev1} for a pedagogical
overview). We eliminate the coordinate ambiguities in the description
of the fluctuations by working in a specific coordinate system, namely
{\it longitudinal gauge}, in which the metric is given by (in the
absence of anisotropic stress)
\be
ds^2 \, = \, a(\tau)^2 \bigl[ (1 + 2 \Phi) d \tau^2 -
(1 - 2 \Phi) d{\bf x}^2 \bigr] \, ,
\ee
where $\tau$ is conformal time related to physical time $t$ via
$dt = a d\tau$, and the relativistic potential $\Phi({\bf x}, \tau)$
is the field describing the fluctuations.

On super-Hubble scales and in an expanding universe, the dominant mode
of $\Phi$ is constant in time if the equation of state of the
cosmological background is constant. We will call this mode the
D-mode. The second fundamental solution, the S-mode, is decreasing in
time. In a contracting phase, the D-mode is sub-dominant, and the
S-mode is growing and hence dominant.

Both in the context of Pre-Big-Bang scenario \cite{PBBflucts} and in
the original effective field theory description of the Ekpyrotic
cosmology \cite{Ekpflucts} it was found that if the fluctuations are
matched across a hypersurface of fixed matter field value (the
hypersurfaces singled out in the scenarios of \cite{PBB,EKP}), then
the growing mode in the contracting universe couples almost
exclusively to the subdominant mode in the expanding phase (the
coupling to the dominant mode is suppressed by $k^2$). In the
Ekpyrotic scenario \cite{EKP} a scale-invariant spectrum of the
dominant mode of cosmological fluctuations is generated in the
contracting phase \cite{KOST}, whereas the decaying mode has a $n = 3$
spectrum \footnote{We are employing the standard notation in which the
index of the spectrum of scalar metric fluctuations is denoted by $n -
1$, with $n = 1$ standing for a scale-invariant spectrum.}. Due to the
suppression of the coupling, only a $n = 3$ spectrum in the expanding
phase is induced \footnote{In the case of the Ekpyrotic scenario, the
underlying physics is higher-dimensional, and the reduction of the
analysis to a four space-time dimensional effective field theory
without adding entropy modes does not correctly model the full physics
of the fluctuations. An analysis of the transfer of metric fluctuations
in five space-time dimensions \cite{Tolley} (see also \cite{Battefeld}
for a similar analysis in a non-singular cosmological background)
shows that in the higher-dimensional setup, an initial scale-invariant
spectrum of cosmological perturbations in the contracting phase goes
over into a scale-invariant spectrum in the expanding phase.}

As initially pointed out in \cite{Peter,Durrer}, the result for the spectrum
of fluctuations in the expanding phase depends quite sensitively on
the details of the matching, and it is in fact not clear whether the
matching prescription of \cite{Hwang,Deruelle} can be applied at all
due to the inconsistency of the matching of the background. It is thus
of great interest to study how the cosmological fluctuations propagate
through a nonsingular cosmological bounce. A first analysis of the
evolution of fluctuations through a specific nonsingular bounce
obtained by using a higher derivative gravity action was made in
\cite{Tsujikawa} (see also \cite{Cartier}), showing that initial
scale-invariant fluctuations do not pass through the bounce, thus
confirming the results of \cite{Ekpflucts}. On the other hand, bounces
obtained by adding spatial curvature and matter with wrong-sign
kinetic terms, but with the standard gravitational action, have been
studied in \cite{Fabio2,Peter,Bozza}, with differing results\footnote{Note that there is a technical problem with some
of the analyses in these references, as discussed in \cite{Thorsten}.}. Whereas
the analysis of \cite{Bozza} yielded no transfer of the growing mode
in the contracting phase to the dominant mode in the expanding period,
the analysis of \cite{Peter} showed that the transfer matrix which
links the two fluctuation modes in the contracting phase to those in
the expanding phase depends quite sensitively on the details of the
bounce, and that it is possible that the coupling of the dominant mode
in the contracting phase to that in the expanding phase is not
suppressed. A similar conclusion was reached recently in a study of
a single field model without spatial curvature in the context of a
theory with modified kinetic term \cite{Abramo} \footnote{See also
\cite{Pinto} for an analysis of how fluctuations arise in a bouncing
cosmology via quantum cosmological methods.}. In the context of a bouncing mirage cosmology it has also been
  shown very recently \cite{BFS} that the spectral index of
   the dominant mode of $\Phi$ does not change.

The approach that we take instead, is to assume that the new physics which is responsible for resolving the singularity does not effect the evolution of the perturbations. The advantage of such an approach is that all the information of the new physics is now encoded in the background behaviour of the Hubble rate, and in particular we can track the evolution of the perturbations unambiguously through the bounce, without having to implement any matching conditions. For some special cases one even has exact analytic expressions which facilitates our understanding. Clearly, such an assumption cannot be valid for arbitrary new physics, but our main motivation for considering it comes from recent progress that has been made in resolving singularities involving ``non-local'' physics. In these cases one can argue that the new physics does not effect the ``local'' perturbation equation but only impacts the ``global'' or background evolution of the universe. For instance, in~\cite{gap} a cosmological BCS theory was considered where fermions can form Cooper pairs and the negative  gap (binding) energy can mediate a bounce. Now the gap energy  depends on the chemical potential of the system which in turn depends on the density of states. As is well known, the density of states in a given system  depends only on the total volume of the system. Thus although the Gap energy contributes to the Hubble equation, it is completely ignorant of local fluctuations of the volume (metric) and therefore does not alter the perturbation equations. This situation is very similar to the usual vacuum energy (cosmological constant)  that gravitates (and therefore contributes to the Hubble equation) but does not contribute to the perturbation equation. Another example of this kind is the Casimir energy, that represents only a global shift in the vacuum energy, which can be negative  and therefore  resolve the singularity without effecting perturbations~\cite{casimir,entropy}. Yet another interesting scenario is to consider higher derivative gravity correction terms to the Einstein action which we expect will become dominant near the bounce, as long as the
curvature at the bounce point becomes comparable to Planck-scale
curvature. Recently, a ghost-free and asymptotically free higher
derivative gravity model leading to a cosmological bounce was proposed
in \cite{Biswas1}, and in a follow-up paper \cite{Biswas2} it was
suggested that a long bouncing phase may lead to the correct thermal
string gas initial conditions for the new structure formation
scenario \cite{NBV,BNPV2} (see also \cite{RHBrev2} for a recent
review). The action of \cite{Biswas1} has the property that, although
the cosmological background is changed, the correction terms in the
equation of motion for long-wavelength (compared to the scale of new physics) cosmological perturbations are
suppressed, and we can thus evolve these perturbations using the usual
equations for cosmological fluctuations~\footnote{Although such a bouncing
cosmology can provide us with a non-singular description of the
universe and possibly a new mechanism for the origin of cosmological
perturbations, the models have not yet addressed the question of why after the bounce, in the expanding branch, we are mostly left with  Standard Model particles. One has to perhaps find a mechanism similar to ``reheating'' where after the bounce most of the energy is converted to Standard Model degrees of freedom. Another possibility could be  that the bounce
creates the initial conditions to drive a phase of late time
inflation \cite{AEGM,AEGJM} and/or create Standard Model baryons and cold
dark matter by exciting the minimal supersymmetric degrees of
freedom~\cite{AEJM} (for a review see \cite{MSSM-REV}).}.

Thus in this paper, we will study the evolution of cosmological
fluctuations using the perturbed Einstein equations through a
specially chosen cosmological bounce background which leads to exactly
soluble equations. In our example, we find that the growing mode in
the contracting phase goes over into the growing mode in the expanding
phase, in contrast to what happens in the model of
\cite{Tsujikawa}. Our result gives rise to the hope that, once the
cosmological singularity in the Ekpyrotic scenario is resolved by a
nonsingular bounce, it will not be necessary to invoke a
scale-invariant spectrum of initial entropy fluctuations
\cite{Notari,Fabio3,Turok2,Ovrut,Senatore,other,BBM} in addition to the
existing scale-invariant spectrum of adiabatic modes in order to
obtain a scale-invariant spectrum in the expanding phase.

\section{The Model}

The starting point of our analysis is the generalization of Einstein's equations to
\be
G_{\mu}^{\nu}=T_{\mu}^{\nu}+Q_{\mu}^{\nu}
\label{gen-GR}
\ee
where $Q_{\mu\nu}$ arises from some uknown new physics responsible for resolving the Big Bang singularity. A priori, $Q_{\mu\nu}$ may arise either in the gravity sector (see for instance~\cite{Biswas1}) or in the matter sector~\cite{casimir,entropy,gap,newwork} but we purposely do not specify it. As usual the above equation can be broken down into the background (which only depends on time) and perturbed (depends on both space and time) quantities  such as
\be
G_{\mu}^{\nu}\equiv <G_{\mu}^{\nu}>(\tau)+\de G_{\mu}^{\nu}(x)
\label{break-up}
\ee
and similarly for $T_{\mu}^{\nu}$ and  $Q_{\mu}^{\nu}$. For homogeneous isotropic cosmology, the background quantities generalize the Hubble equation to
\be
3H^2=<G_{0}^{0}>=<T_{0}^{0}>+<Q_{0}^{0}>=<\rho>+<Q_{0}^{0}>
\ee

To keep things as general as possible let us not specify the new
physics, \ie  $<Q_{0}^{0}>$ that causes the universe to bounce, but simply make an ansatz\footnote{If one wants, one can deduce $<Q_{0}^{0}>$ from the ansatz.}
for the time evolution of the Hubble rate. Near the bounce, the most generic behavior of the scale factor is given by
\be
a(\tau)=1+\LF{\tau\over\tau_0}\RF^2
\label{bounce-a}
\ee
where $\tau_0$ corresponds to the bounce time scale. (\ref{bounce-a}) leads to a Hubble evolution of the form
\be \label{form}
\cH \, \sim\, {\tau\over \tau^2+\tau_0^2} \, ,
\ee
where $\cH$ is the Hubble rate in conformal time. Now, we also know
that at late times $\cH$ has to
have the right asymptotic property, $\cH\ra {q\over \tau}$, with $q<1$. Of course, in practice the precise transition from the accelerating bouncing region (\ref{form}) to the deccelerating late time regime will depend on the new physics that is introduced. However, at least in the case where there is really only one new fundamental scale we expect only the details to depend on the precise functional form of the transition. Thus, for the purpose of illustration and technical simplicity we choose to work with a specific ansatz\footnote{If we believe that the
singularity at $\tau=0$ will be resolved, then it is clear that the
pole at $\tau=0$ will have to shift from the real axis to the complex
plane. Since we want $\cH$ to be finite and real in the entire real
axis, it is easy to see that in fact the poles must lie along the
imaginary axis and come in conjugate pairs. Therefore, the simplest
ansatz that one can make for $\cH$, is to consider only a pair of
simple poles as in (\ref{ansatz}). However, the general algorithm that will be advocated here can be carried forward for any non-singular bounce ansatz that one may get for a given new physics.
Also, we believe that this ansatz for the Hubble parameter may be realized as non perturbative solutions of non metric theories of gravity as well as from four dimensional, higher order corrected Heterotic M theory actions.  These constructions will be pursued in future work \cite{newwork}. } to capture the
essential effects on the perturbations as they evolve across the
bounce:
\be \label{ansatz}
\cH \, = \, {q\tau\over \tau^2+\tau_0^2} \, ,
\ee
We note that this has the right asymptotic property as
$\tau\ra\pm\infty$, while during the bounce $\cH$ goes linearly with
$\tau$ as expected.

As matter we consider a fluid with an equation of state
\be
p \, = \, \om \rho \, ,
\ee
where $\rho$ and $p$ stand for the energy density and the pressure
of the fluid, respectively. Inserting this equation of state into
the general relativistic Friedmann equations yields the following
time evolution of the scale factor:
\be \label{p}
a(\tau)=a_0|\tau|^q\mx{ with }q\equiv \frac{2}{1+3\om}\, ,
\ee
where $a_0$ is a constant.

\section{Perturbations across the Bounce}
Starting from (\ref{gen-GR}) and using perturbative expansions such as (\ref{break-up}) we obtain the generalized perturbation equation:
\be
\de G_{\mu}^{\nu}=\de T_{\mu}^{\nu}+\de Q_{\mu}^{\nu}
\label{gen-pert}
\ee
We now make the crucial assumption that $  \de Q_{\mu}^{\nu}$ vanishes (or remains negligible during the bounce). As mentioned in the Introduction, when the resolution of the Big Bang singularity involves new non-local physics such as~\cite{gap,casimir,entropy,Biswas1,Biswas2}, then at least for fluctuations whose length scales during the bounce are much larger than $\tau_0$, this can be justified.

Returning to (\ref{gen-pert}), one can now check that just by using the usual expressions of $\de   G_{\mu}^{\nu}$ and $\de   T_{\mu}^{\nu}$ in terms of $\Phi$ and $\de \rho$  for an ideal gas, one can derive the usual General Relativistic perturbation equation for the Bardeen potential $\Phi$
\be
\Phi_k^{''} + 3(1+\om)\cH\Phi_k^{'} + \om k^2\Phi_k +
[2\cH^{'}+(1+3\om)\cH^2]\Phi_k \, = \, 0 \, .
\label{gen-Phi}
\ee
Crucially we note that this derivation does not ``re-use'' the background Hubble equation and therefore goes through even though the background equation is modified. As we have expalained in the appendix, this is not the case for the standard evolution equation for the Mukhanov variable (also see section \ref{sec;discussion}) and in fact this is the reason why one cannot trust the General Relativistic equation for the Mukhanov variable  to track perturbations around the bounce.

In passing we note that for GR, in case of an ideal fluid the last term
drops out. However, when one modifies gravity, this is no longer true,
and this is what makes the analysis a lot more interesting.

Substituting our ansatz (\ref{ansatz}) into (\ref{gen-Phi}),
we  obtain the following
equation for the metric fluctuations
\be
\Phi_k^{''} + {3(1+\om)q\tau\over\tau^2+\tau_0^2}\Phi_k^{'} +
\LT\om k^2+{2q\tau_0^2\over (\tau^2+\tau_0^2)^2}\RT\Phi_k \, = \, 0 \, .
\label{bounce-Phi}
\ee
As one can see, the coefficients in the above differential equation are all non-singular and
completely well defined. Therefore one can in principle solve for
$\Phi_k$ in the interval $(-\infty,\infty)$ and understand how an initial
perturbation evolves from the contracting to the expanding phase, without
having to implement any matching conditions anywhere. Although this is in
general only possible to achieve numerically, as we will argue now, one
can make significant progress analytically.

Let us first identify some of the regimes where the differential equation
may simplify. Firstly, one has the usual division between the super- and
sub-Hubble regimes separated at points where
\be
k \, = \, |\cH| \,\,\, \Ra \,\,\, k|\tau|^2-q|\tau|+\tau_0^2k \,
= \, 0 \, .
\ee
As is typical in bouncing cosmologies, there are four solutions
to the above equation:
\be
\tau \, = \,
\pm\LF{q\pm\sqrt{q^2-4k\tau_0}\over 2k}\RF\equiv \pm\tau_{\pm} \, .
\ee
Two occur during contraction ($-\tau_{\pm}$), and two during expansion
($\tau_{\pm}$). The initial and final crossings are the usual ones for
a contracting and an expanding universe, respectively, whereas the two
middle ones occur near the bounce point and are a special feature of
nonsingular bouncing cosmologies. The first crossing occurs at
$-\tau_+$, when GR is a valid description of the background and modes
exit the Hubble radius (note the universe starts from a cold $\cH\ra
0$ phase, so that all the modes are sub-Hubble to begin with). Around
$\tau\sim\tau_0$, the new physics kicks in and $\cH$ starts to
decrease again. Thus, all the modes which are super-Hubble start to
come back inside the Hubble radius, and for a given mode this happens
at $-\tau_-$. By the time the bounce occurs all the modes are again
sub-Hubble, but after the bounce as the Hubble radius reaches a finite
value, modes again exit the Hubble radius. This happens at
$\tau_-$. Finally, as the expansion continues according to the
equations of standard cosmology, modes eventually enter our horizon at
$\tau_+$.

A second way to divide the evolution is to consider the range
$|\tau|>\tau_0$ and $|\tau|<\tau_0$ separately, where the former
corresponds to considering the intervals when GR is a good description
of the background, while for the latter one has to include the effects
of the new physics.  So, how should one proceed?

First let us look at (\ref{bounce-Phi}) as $|\tau|\gg \tau_0$, where
we recover the general relativistic limit
\be
\Phi_k''+{6(1+\om)\over \tau(1+3\om)}\Phi_k'+\om k^2\Phi_k \, = \, 0 \, ,
\label{Phi-GR}
\ee
which has the following analytical solution
\be
\Phi_k \, = \, \tau^{-\nu}
[k^{-\nu}D_{\pm}(k)J_{\nu}(\sqrt{\om}k\tau)+k^{\nu}S_{\pm}(k)J_{-\nu}(\sqrt{\om}k\tau)] \, ,
\label{phi-sol}
\ee
where
\be
\nu \, = \, {1\over 2}\LF{5+3\om\over 1+3\om}\RF \, .
\ee
and $+$ or $-$ labels the coefficients corresponding to the expanding
$(\tau_0,\infty)$ and contracting $(-\infty,-\tau_0)$ phases
respectively where the solution is a good approximation. In
particular, it incorporates the sub-Hubble phases at early and late
times, \ie between $(-\infty,-\tau_+)$ and $(\tau_+,\infty)$. Also, we have included the $k$-dependent factors in front of the coefficients $D_{\pm}$ and $S_{\pm}$ to ensure that in the super-Hubble phase the $k$ dependence is completely contained in $D(k)$ and $S(k)$. By using the asymptotic form of Bessel functions:
\be
\lim_{x\ra 0}J_{\nu}(x) \, = \, x^{\nu}\,,
\ee
and  taking the appropriate $\tau\ra 0$ limit of (\ref{phi-sol}), this can easily be seen:
\be
\lim_{\tau\ra 0}\Phi_k \, =\, D_{\pm}(k)\om^{\nu/2}+S_{\pm}(k)\om^{-\nu/2}(\pm\tau)^{-2\nu}\,.
\label{Phi-super}
\ee

Now of
course, if we know the solution to (\ref{bounce-Phi}) in the entire range
$(-\infty,\infty)$, then we will know how $\{D_+,S_+\}$ is related to
$\{D_-,S_-\}$, but as we will now explain, one requires less. Finding
this ``transfer matrix'' without any ambiguities is in fact the main
endeavor of this paper.

In order to achieve this let us next look at the super-Hubble phases. As usual, in these
phases, the $k$-terms can be ignored, as $\cH>k$ and the second term
dominates.  Finally, we are left with the sub-Hubble phase around the
bounce and now comes a crucial observation. Unlike in the usual GR
scenario, in this phase which occurs once the new physics has kicked
in, \ie $|\tau|<\tau_0$, the $k$-term can again be ignored in favor
of the fourth term. This is because for the modes that we observe
today, typically\footnote{If there is no prolonged inflationary
phase.}  $k\ll\tau_0^{-1}$, the latter being the scale of new physics which is
expected to be close to the string or the Planck scale. As a result,
we find that for the entire range $(-\tau_+,\tau_+)$ we can ignore the
$k$-term and need only solve
\be
\Phi_k'' + {3(1+\om)q\tau\over\tau^2+\tau_0^2}\Phi_k'
+ {2q\tau_0^2\over (\tau^2+\tau_0^2)^2}\Phi_k \, = \, 0 \, .
\label{pert-bounce}
\ee

A few comments are now in order. Firstly, observe that we have
completely bypassed having to do any matching at $\pm\tau_-$ or at
$\tau=0$, as is often done in literature. Secondly, the formalism
described here divides the entire evolution into three {\it
overlapping} regions: $(-\infty,-\tau_0)$, $(-\tau_+,\tau_+)$ and
$(\tau_0,\infty)$.  As a result, the matching of the solutions to
(\ref{Phi-GR}) and (\ref{pert-bounce}) can be done unambiguously by
looking at the asymptotic properties of the solutions. This is to be
contrasted with matching conditions which are imposed at specific
points in time (like between sub- and super-Hubble phases at the point
of Hubble crossing). A final remark, as is clear the sub-Hubble phase
around the bounce is very different from the usual sub-Hubble phases
(for instance in inflationary cosmology) because the $k$-term is
unimportant, and thus any analysis based on the usual framework of
solving the equation keeping only the $k$-term is bound to give
incorrect results.

Above, we already obtained exact solutions (\ref{phi-sol}) for the Ist
and IIIrd region, \ie in the intervals $(-\infty,-\tau_0)$ and
$(\tau_0,\infty)$. In order to avoid any ambiguity involving mode
matching we need to find exact solutions to (\ref{pert-bounce}). To do
this it is convenient to define
\be
y_k \, \equiv \, \Phi_k(\tau^2+\tau_0^2)^{-n} \, .
\label{y}
\ee
The differential equation then becomes
\bea \label{yeom}
&&(\tau^2+\tau_0^2)^2y'' +
(\tau^2+\tau_0^2)\tau(4n+3(1+\om)q)y' \nonumber \\
&+&
[2n(\tau^2+\tau_0^2)+(4n(n-1) \\
&& + 6nq(1+\om))\tau^2+2q\tau_0^2]y \,
= \, 0 \, \nonumber .
\eea
If we now choose $n$ to satisfy
\be
4n(n-1)+6nq(1+\om) \, = \, 2q \, ,
\label{n}
\ee
which leads to
\be
n_{\pm} \, = \, {-1\pm\sqrt{2+3\om}\over 1+3\om} \, ,
\ee
and in particular choose to work with $n=n_-$, then the ordinary
differential equation greatly simplifies and one has
\be
(\tau^2+\tau_0^2)y'' + \tau\bb y' - 2n_+y \, = \, 0 \, ,
\ee
where
\be
\bb\equiv {2(1+3\om-\sqrt{2+3\om})\over 1+3\om}
\ee

\section{Mode Switch: an Exact Special Case}

In order to have a clear understanding of the physics we will focus on
a special case where the solution can be written in terms of familiar
functions. This is the case when the coefficient of the second term,
$\bb$, equals one, which happens when $\om\approx 5.314$.

We note in passing that such a large $\om$ can actually be physically
interesting as it is known that as $\om\ra\infty$ we produce a
scale-invariant spectrum in one of the modes (the Ekpyrotic scenario).
In any case, the reason we want to consider the above $\om$ is because
it lends technical simplicity to extract the physics which will be
useful when we study the more general case.

For  the above special case the differential equation simplifies to
\be
(\tau^2+\tau_0^2)y''+\tau y'-\al^2 y \, = \, 0 \,
\ee
with
\be
\al^2 \, \equiv \, 2n_+ \, > \, 0 \, .
\ee
Its solution reads
\be
y \, = \, B_1\exp\LF\al\sinh^{-1}{\tau\over\tau_0}\RF
+ B_2\exp\LF-\al\sinh^{-1}{\tau\over\tau_0}\RF \, .
\label{y-sol}
\ee
In order to understand how the perturbations propagate through the
bounce we have to look at the behavior of $y$ as $\tau\ra\pm\infty$.
Now,
\be
\lim_{\tau\ra\pm\infty}\sinh^{-1}\LF{\tau\over\tau_0}\RF \,
= \, \pm\ln\LF{\pm 2\tau\over\tau_0}\RF \, .
\ee
As a result, the asymptotic values of $y$ reads
\be
\lim_{\tau\ra\pm\infty}y \, = \,
B_1\LF{\pm 2\tau\over\tau_0}\RF^{\pm\al} +
B_2\LF{\pm 2\tau\over\tau_0}\RF^{\mp\al} \, .
\ee

Something remarkable has happened: in going from contraction to
expansion, the coefficients corresponding to the two modes
$\tau^{\al}$ and $\tau^{-\al}$ have completely switched. The dominant
mode in the contracting phase goes over into the dominant mode in the
expanding phase, unlike what happens in a singular bounce making use
of the Hwang-Vishniac and Deruelle-Mukhanov matching conditions. To see this more precisely, by matching the bouncing solution (\ref{y-sol}) to the late time solutions given by (\ref{phi-sol}) in the overlapping regions one finds the exact relation:
\be
D_+=S_-\LF{\om\tau_0^2\over 2}\RF^{-\nu}\mx{ and }S_+=D_-\LF{\om\tau_0^2\over 2}\RF^{\nu}
\ee

For a
general equation of state, although we do not expect such a complete
switch, the above result certainly suggests that there would be mode
mixing.

\section{Mode Mixing as  $\om\gg 1$}

It is known that the Ekpyrotic scalar which mimics an ideal fluid with
a large equation of state parameter produces a (nearly)
scale-invariant spectrum of perturbations during the phase of contraction,
but in the growing mode. It was argued that this mode matches
to the decaying mode exclusively during the expansion phase and
therefore cannot explain the near scale-invariance observed in the CMB
today. The previous exact analysis already suggests that this may not
be true if the Ekpyrotic bounce is smoothed out.

Quite remarkably, in the $\om\ra\infty$ limit the
differential equation (\ref{yeom}) simplifies to
\be (\tau^2+\tau_0^2)y'' +
2\tau y' \, = \, 0 \, , \ee
and again becomes amenable to an exact treatment. The equation
has the following solution
\be
y \, = \, B_1+B_2\tan^{-1}(\tau/\tau_0) \,.
\label{ekpyrotic}
\ee
Now, in the case $\om\gg 1$ we obtain from (\ref{Phi-super})
\be
y_k \, \approx \, \Phi_k \, = \, D_{\pm}(k)\om^{1/4}+S_{\pm}(k)
\om^{-1/4}(\pm\tau)^{-1} \,,
\ee
where we have used that $n\ra 0$ as $\om\ra\infty$. We can now relate
the $+$ coefficients with $-$ via the bounce solution
(\ref{ekpyrotic}). By considering the $\tau\ra\pm\infty$ limit of
(\ref{ekpyrotic}) one easily finds
\be \om^{1/4}
D_{\pm} \, = \, B_1\pm{\pi B_2\tau_0\over 2}\mx{ and
}\pm\om^{-1/4}S_{\pm}=-B_2\tau_0\,. \ee
This leads us to
\be
D_+(k) \, = \, D_-(k)-{\pi \over\om^{1/4} \tau_0}S_-(k)\mx{ and }S_+(k)=-S_-(k) \,.\ee
We notice that the constant mode $D_+$ gets contributions from both the
modes in the contracting phase. In particular, the contribution
from $S_-$ gives a scale invariant spectrum \footnote{To see this note that according to conventional quantization of the Mukhanov variable, as $\tau\ra-\infty$, $v\sim (V_0/ \sqrt{k})[\cos(\sqrt{\om} k\tau)+\sin(\sqrt{\om} k\tau)]$. Using the relation $v\sim \tau^{3(1+\om)/( 1+3\om)}[\Phi'+2\nu(\Phi/\tau)]$ valid for ideal fluids in the General Relativistic limit $(-\infty,-\tau_+)$, one then finds $D_-\sim k^{-1/2}$ and $S_{-}\sim k^{-3/2}$, the latter giving rise to a scale-invariant contribution to the power spectrum.}!

\section{Discussion}\label{sec:discussion}

In this paper, we have followed the evolution of cosmological
fluctuations across the bounce using the relativistic potential
$\Phi$, the metric fluctuation in longitudinal gauge. From experience
built up in investigations of inflationary universe models (see
e.g. \cite{BK,BST}), Pre-Big-Bang cosmologies and the Ekpyrotic
scenario, working in terms of $\Phi$ can be dangerous because of very
sensitive dependence of the evolution on matching conditions at times
when the equation of state of the cosmological background changes,
e.g. during the period of reheating in the context of inflationary
cosmology. In our example we are safe from this danger because we have
an exact solution and hence do not need to invoke any matching
conditions.

On the other hand, from the point of view of the quantum theory of
cosmological perturbations (see \cite{Sasaki,Mukhanov} for pioneering
works), a variable different from $\Phi$ carries more physical
meaning, namely the variable $v$ in terms of which the action for
cosmological fluctuations has canonical kinetic term. In the case of
hydrodynamical matter, the variable $v$, which determines the
curvature perturbation in comoving gauge, is related to $\Phi$ via
\be
c_s v \, \equiv \, u^{'} - \bigl({{\theta^{'}} \over {\theta}} \bigr) u
\label{v-defn}
\, ,
\ee
where $c_s$ is the speed of sound,
\be
\te \, \equiv \, {1\over a}\LF 1+{p\over\rho}\RF^{-1/2} \, ,
\ee
and
\be
u \, \equiv \, {\Phi\over \sqrt{\rho+p}} \, .
\label{u-defn}
\ee

The variable $v$ obeys the following field equation:
\be
v''-c_s^2\n^2v-{z''\over z}v \, = \, 0 \, ,
\label{v-eqn}
\ee
where
\be
z \, \equiv \, {a\sqrt{\bb}\over -\cH c_s}\mx{ and }
\bb \, \equiv \, \cH^2-\cH' \, .
\ee
In inflationary cosmology, the variable $v$ remains constant during
the transition of the equation of state which takes place at the time
of reheating, whereas $\Phi$ jumps by a large factor. It is thus the
variable $v$ which is a more robust one to follow. In fact, in the
absence of entropy fluctuations, one can show that on super-Hubble
scales, the variable
\be
{\cal R} \, = \, {v \over z}
\ee
is conserved \cite{BST,BK,Lyth}:
\be
(1 + \om) {\dot{\cal R}} \, = \, 0 \, .
\label{conservation}
\ee
In the contracting phase of an Einstein universe it can be shown
\cite{Ekpflucts} that the dominant mode of $\Phi$ does not couple to
the variable $v$, and hence the curvature fluctuation is only
sensitive to the decaying mode of $\Phi$, a mode which does not lead
to a scale-invariant spectrum.

In a bouncing cosmology, it becomes problematic to use the equation of
motion (\ref{v-eqn}) for $v$. For instance, at the bounce point the variable $z$ blows up
and hence the equation is singular. Essentially the new physics which solves the Big Bang singularity necessarily has to modify (\ref{v-eqn}). As we discuss in details in the appendix, in general it is not even possible to come up with an equation for $v$ (see~\cite{Pinho} for an attempt in this direction) while at least in our scenario the perturbation equation for $\Phi$ is still correct. Thus it is not justified to use the
large scale analysis and conclude that ${\cal R}$ is constant.  These
are our reasons for focusing on the evolution equation for
$\Phi$ and not for $v$.

\section{Conclusions}

In this paper we have presented a particular nonsingular bouncing
cosmological background, in which the usual general relativistic
equations for cosmological perturbations can be solved exactly. We
find that the growing mode in the contracting phase goes over into the
dominant mode in the expanding phase. This is unlike what happens in
four dimensional effective field theories of Pre-Big-Bang or Ekpyrotic
type, in which the fluctuations are matched at a distinguished but
singular hypersurface using the analog of the Israel matching
conditions. Our result supports the conclusions of \cite{Durrer,Peter}
that the transfer of fluctuations is very sensitive to the details of
the bounce.

Our analysis assumes that terms in the Lagrangian different from the
Einstein-Hilbert terms generate the nonsingular bounce. At the same
time, it is crucial that these new terms not effect the equations for
the IR fluctuation modes. An example where both conditions are
satisfied is given in \cite{Biswas1,Biswas2}.

Applied to the Ekpyrotic scenario, our result implies that one may not
be required to invoke entropy fluctuations with a scale-invariant
spectrum \cite{Notari,Fabio3,Turok2,Ovrut,Senatore}
in order to obtain a scale-invariant spectrum of curvature fluctuations
in the expanding phase.

\begin{acknowledgments}
We especially want to thank Anupam Mazumdar for collaborating on this project in the initial stages.  SA and TB are supported by the Eberly College of Science at Penn State, the work of RB is supported by an NSERC Discovery Grant, by the
Canada Research Chairs Program, and by funds from a FQRNT Team
Grant.

\end{acknowledgments}
\section{Appendix}

In this appendix we explain why in our bouncing universe scenario we used the Bardeen potential $\Phi$ to track the perturbations across the bounce rather than the more conventional Mukhanov variable $v$. The discussion will also clarify how we can evade the usual conservation arguements involving the curvature perturbation (\ref{conservation}) to get mode-mixing.
Let us first define the intermediate $u$ variable:
\be
u\equiv \exp\LT{3\over2}\int(1+c_s^2)\caH\ d\tau\RT\Phi
\ee
For an ideal gas using the continuity equation
\be
\rho'+3\caH(\rho+p)=0
\ee
$u$ simplifies to
\be
u={\Phi\over \sqrt{\rho}}
\ee
If one now uses the relation
\be
2\caH'+(1+3\om)\caH^2=0
\label{Hrelation}
\ee
which is valid only for a General Relativistic background one can derive a relatively simple differential equation for $u$ from (\ref{gen-Phi}):
\be
u''-\om\n^2 u-{\te''\over \te}u=0\mx{ where }\te\equiv {1\over a\sqrt{1+\om}}
\label{u-eqn}
\ee

Since the new physics that resolves the singularity alters the background equations, (\ref{Hrelation}) can no longer be valid (in fact one can check that (\ref{Hrelation}) always lead to a singular universe), and therefore one cannot trust (\ref{u-eqn}). Nevertheless, one can actually derive a similar equation for $u$ which does not use (\ref{Hrelation}) and therefore will is valid irrespective of the new physics. The equation reads:
\be
u''-\om\n^2 u-(\al\caH'+\bb\caH^2)u=0
\label{u-gen}
\ee
where
\be
\al\equiv {3\om-1\over 2}\mx{ and } \bb\equiv{9\om^2+6\om-5\over 4}
\ee
One can easily verify that provided (\ref{Hrelation}) is valid, (\ref{u-gen}) reduces to (\ref{u-eqn}).

In the usual General Relativistic analysis one typically defines the Mukhanov variable via (\ref{v-defn}) which for an ideal gas simplifies to
\be
v=u'+\caH u
\ee
and turns out to be the right canonical variable for the effective quantum action~\cite{MFB,RHBrev1} (again in the General Relativistic limit) and therefore let's one impose the ``sub-Hubble  quantum fluctuations'' as initial conditions. In this context we note that using  the relations between $\Phi,u$ and $v$ (\ref{u-defn},\ref{v-defn}), one can also read off the appropriate initial conditions for $\Phi$ or $u$, from that of $v$, and then choose to analyse the propagation of the fluctuation in any of the variables that one chooses to (see footnote 9 for instance). We should also point out that all the definitions (\ref{u-defn},\ref{v-defn}) are non-singular, and one expects the solutions for all these variables to be regular when one uses the correct evolution equations for them.

Actually, this is where the Mukhanov variable as defined in (\ref{v-defn}) looses it's usefulness. Unlike $\phi$ and $u$, one cannot even find an evolution  equation  for $v$ like (\ref{v-eqn}) in the general case when one has introduced some new physics to resolve the singularity. Instead one finds the following equation
\be
v''-\om\n^2v-[(\al+2)\caH'+\bb\caH^2](u'+f u)=0
\ee
where we have defined a function which depends on the background:
\be
f\equiv{\bb\caH^3+(\al+1)\caH''+(2\bb+\al)\caH\caH'\over(\al+2)\caH'+\bb\caH^2}
\ee
Clearly, in order to have an equation only in terms of $v$, the terms involving $u$ must combine to give us $v$ which can happen  iff one can satisfy
\be
3(1+\om)\caH''+(1+3\om)^2\caH\caH'=0
\ee
Indeed for GR this condition is satisfied. However, it is clear that when one introduces new physics, in general there is no simple evolution equation for $v$ and (\ref{v-eqn}) is certainly not valid\footnote{It may be an interesting exercise to check whether it is possible to generalize the definition of $v$ such that it reduces to (\ref{v-eqn}) in the  GR limit, but differs from it near the bounce in such a way that one can obtain an evolution equation for $v$ like it's GR cousin. For such an attempt in the context of quantum cosmological models see~\cite{Pinho}.}. It also explains why one can't use the conservation law (\ref{conservation}) which is based on (\ref{v-eqn}). This is why a mode-mixing is actually possible in our scenario.


\begin{thebibliography}{99}

\bibitem{Hawking}
S.~W.~Hawking and R.~Penrose,
  ``The Singularities of gravitational collapse and cosmology,''
  Proc.\ Roy.\ Soc.\ Lond.\  A {\bf 314}, 529 (1970).

\bibitem{Borde}
A.~Borde and A.~Vilenkin,
  ``Eternal inflation and the initial singularity,''
  Phys.\ Rev.\ Lett.\  {\bf 72}, 3305 (1994)
  [arXiv:gr-qc/9312022].

\bibitem{PBB}
M.~Gasperini and G.~Veneziano,
  ``Pre - big bang in string cosmology,''
  Astropart.\ Phys.\  {\bf 1}, 317 (1993)
  [arXiv:hep-th/9211021].;\\
M.~Gasperini and G.~Veneziano,
  ``The pre-big bang scenario in string cosmology,''
  Phys.\ Rept.\  {\bf 373}, 1 (2003)
  [arXiv:hep-th/0207130];\\
J.~E.~Lidsey, D.~Wands and E.~J.~Copeland,
  ``Superstring cosmology,''
  Phys.\ Rept.\  {\bf 337}, 343 (2000)
  [arXiv:hep-th/9909061].

\bibitem{EKP}
J.~Khoury, B.~A.~Ovrut, P.~J.~Steinhardt and N.~Turok,
  ``The ekpyrotic universe: Colliding branes and the origin of the hot big
  bang,''
  Phys.\ Rev.\  D {\bf 64}, 123522 (2001)
  [arXiv:hep-th/0103239].

\bibitem{Biswas1}
T.~Biswas, A.~Mazumdar and W.~Siegel,
  ``Bouncing universes in string-inspired gravity,''
  JCAP {\bf 0603}, 009 (2006)
  [arXiv:hep-th/0508194].

\bibitem{Israel}
W.~Israel,
  ``Singular hypersurfaces and thin shells in general relativity,''
  Nuovo Cim.\  B {\bf 44S10}, 1 (1966)
  [Erratum-ibid.\  B {\bf 48}, 463 (1967\ NUCIA,B44,1.1966)].

\bibitem{Hwang}
J.~c.~Hwang and E.~T.~Vishniac,
  ``Gauge-invariant joining conditions for cosmological perturbations,''
  Astrophys.\ J.\  {\bf 382}, 363 (1991).

\bibitem{Deruelle}
N.~Deruelle and V.~F.~Mukhanov,
  ``On matching conditions for cosmological perturbations,''
  Phys.\ Rev.\  D {\bf 52}, 5549 (1995)
  [arXiv:gr-qc/9503050].

\bibitem{MFB}
V.~F.~Mukhanov, H.~A.~Feldman and R.~H.~Brandenberger,
  ``Theory of cosmological perturbations. Part 1. Classical perturbations. Part
  2. Quantum theory of perturbations. Part 3. Extensions,''
  Phys.\ Rept.\  {\bf 215}, 203 (1992).

\bibitem{RHBrev1}
R.~H.~Brandenberger,
  ``Lectures on the theory of cosmological perturbations,''
  Lect.\ Notes Phys.\  {\bf 646}, 127 (2004)
  [arXiv:hep-th/0306071].

\bibitem{PBBflucts}
R.~Brustein, M.~Gasperini, M.~Giovannini, V.~F.~Mukhanov and G.~Veneziano,
  ``Metric perturbations in dilaton driven inflation,''
  Phys.\ Rev.\  D {\bf 51}, 6744 (1995)
  [arXiv:hep-th/9501066].

\bibitem{Ekpflucts}
D.~H.~Lyth,
  ``The primordial curvature perturbation in the ekpyrotic universe,''
  Phys.\ Lett.\  B {\bf 524}, 1 (2002)
  [arXiv:hep-ph/0106153];\\
D.~H.~Lyth,
  ``The failure of cosmological perturbation theory in the new ekpyrotic
  scenario,''
  Phys.\ Lett.\  B {\bf 526}, 173 (2002)
  [arXiv:hep-ph/0110007];\\
R.~Brandenberger and F.~Finelli,
  ``On the spectrum of fluctuations in an effective field theory of the
  ekpyrotic universe,''
  JHEP {\bf 0111}, 056 (2001)
  [arXiv:hep-th/0109004];\\
J.~c.~Hwang,
  ``Cosmological structure problem in the ekpyrotic scenario,''
  Phys.\ Rev.\  D {\bf 65}, 063514 (2002)
  [arXiv:astro-ph/0109045];\\
J.~Khoury, B.~A.~Ovrut, N.~Seiberg, P.~J.~Steinhardt and N.~Turok,
  ``From big crunch to big bang,''
  Phys.\ Rev.\  D {\bf 65}, 086007 (2002)
  [arXiv:hep-th/0108187];\\
P.~Creminelli, A.~Nicolis and M.~Zaldarriaga,
  ``Perturbations in bouncing cosmologies: Dynamical attractor vs scale
  invariance,''
  Phys.\ Rev.\  D {\bf 71}, 063505 (2005)
  [arXiv:hep-th/0411270].

\bibitem{KOST}
J.~Khoury, B.~A.~Ovrut, P.~J.~Steinhardt and N.~Turok,
  ``Density perturbations in the ekpyrotic scenario,''
  Phys.\ Rev.\  D {\bf 66}, 046005 (2002)
  [arXiv:hep-th/0109050].

\bibitem{Tolley}
A.~J.~Tolley, N.~Turok and P.~J.~Steinhardt,
  ``Cosmological perturbations in a big crunch / big bang space-time,''
  Phys.\ Rev.\  D {\bf 69}, 106005 (2004)
  [arXiv:hep-th/0306109].

\bibitem{Battefeld}
T.~J.~Battefeld, S.~P.~Patil and R.~H.~Brandenberger,
  ``On the transfer of metric fluctuations when extra dimensions bounce or
  stabilize,''
  Phys.\ Rev.\  D {\bf 73}, 086002 (2006)
  [arXiv:hep-th/0509043].

\bibitem{Peter}
J.~Martin, P.~Peter, N.~Pinto Neto and D.~J.~Schwarz,
  ``Passing through the bounce in the ekpyrotic models,''
  Phys.\ Rev.\  D {\bf 65}, 123513 (2002)
  [arXiv:hep-th/0112128];\\
P.~Peter and N.~Pinto-Neto,
  ``Primordial perturbations in a non singular bouncing universe model,''
  Phys.\ Rev.\  D {\bf 66}, 063509 (2002)
  [arXiv:hep-th/0203013];\\
P.~Peter, N.~Pinto-Neto and D.~A.~Gonzalez,
  ``Adiabatic and entropy perturbations propagation in a bouncing universe,''
  JCAP {\bf 0312}, 003 (2003)
  [arXiv:hep-th/0306005];\\
J.~Martin and P.~Peter,
  ``On the 'causality argument' in bouncing cosmologies,''
  Phys.\ Rev.\ Lett.\  {\bf 92}, 061301 (2004)
  [arXiv:astro-ph/0312488];\\
J.~Martin and P.~Peter,
  ``Parametric amplification of metric fluctuations through a bouncing
  phase,''
  Phys.\ Rev.\  D {\bf 68}, 103517 (2003)
  [arXiv:hep-th/0307077];\\
J.~Martin and P.~Peter,
  ``On the properties of the transition matrix in bouncing cosmologies,''
  Phys.\ Rev.\  D {\bf 69}, 107301 (2004)
  [arXiv:hep-th/0403173].

\bibitem{Durrer}
R.~Durrer and F.~Vernizzi,
  ``Adiabatic perturbations in pre big bang models: Matching conditions and
  scale invariance,''
  Phys.\ Rev.\  D {\bf 66}, 083503 (2002)
  [arXiv:hep-ph/0203275].

\bibitem{Tsujikawa}
S.~Tsujikawa, R.~Brandenberger and F.~Finelli,
  ``On the construction of nonsingular pre-big-bang and ekpyrotic cosmologies
  and the resulting density perturbations,''
  Phys.\ Rev.\  D {\bf 66}, 083513 (2002)
  [arXiv:hep-th/0207228].

\bibitem{Cartier}
C.~Cartier, J.~c.~Hwang and E.~J.~Copeland,
  ``Evolution of cosmological perturbations in non-singular string
  cosmologies,''
  Phys.\ Rev.\  D {\bf 64}, 103504 (2001)
  [arXiv:astro-ph/0106197];
  S.~Kawai and J.~Soda,
  Phys.\ Lett.\  B {\bf 460}, 41 (1999)
  [arXiv:gr-qc/9903017].

\bibitem{Fabio2}
F.~Finelli,
  ``Study of a class of four dimensional nonsingular cosmological bounces,''
  JCAP {\bf 0310}, 011 (2003)
  [arXiv:hep-th/0307068];\\
L.~E.~Allen and D.~Wands,
  ``Cosmological perturbations through a simple bounce,''
  Phys.\ Rev.\  D {\bf 70}, 063515 (2004)
  [arXiv:astro-ph/0404441].


\bibitem{Bozza}
 V.~Bozza,
  JCAP {\bf 0602}, 009 (2006)
  [arXiv:hep-th/0512066];
V.~Bozza and G.~Veneziano,
  ``Scalar perturbations in regular two-component bouncing cosmologies,''
  Phys.\ Lett.\  B {\bf 625}, 177 (2005)
  [arXiv:hep-th/0502047];\\
V.~Bozza and G.~Veneziano,
  ``Regular two-component bouncing cosmologies and perturbations therein,''
  JCAP {\bf 0509}, 007 (2005)
  [arXiv:gr-qc/0506040].
 M.~Gasperini, M.~Giovannini and G.~Veneziano,
  ``Perturbations in a non-singular bouncing universe,''
  Phys.\ Lett.\  B {\bf 569}, 113 (2003)
  [arXiv:hep-th/0306113].

\bibitem{Thorsten}
T.~J.~Battefeld and G.~Geshnizjani,
  ``A note on perturbations during a regular bounce,''
  Phys.\ Rev.\  D {\bf 73}, 048501 (2006)
  [arXiv:hep-th/0506139].

\bibitem{Abramo}
L.~R.~Abramo and P.~Peter,
  ``K-Bounce,''
  arXiv:0705.2893 [astro-ph].

\bibitem{Pinto}
P.~Peter, E.~J.~C.~Pinho and N.~Pinto-Neto,
  ``A non inflationary model with scale invariant cosmological
  perturbations,''
  Phys.\ Rev.\  D {\bf 75}, 023516 (2007)
  [arXiv:hep-th/0610205].

\bibitem{BFS}
R. Brandenberger, H. Firouzjahi and O. Saremi,
  ``Cosmological perturbations on a bouncing brane,"
  arXiv:0707.4181 [hep-th].



\bibitem{gap}
S.~Alexander and T.~Biswas,
  arXiv:0807.4468 [hep-th].

\bibitem{casimir}
  C.~A.~R.~Herdeiro and M.~Sampaio,
  ``Casimir energy and a cosmological bounce,''
  Class.\ Quant.\ Grav.\  {\bf 23}, 473 (2006)
  [arXiv:hep-th/0510052].
\bibitem{entropy}
  T.~Biswas,
  arXiv:0801.1315 [hep-th].

\bibitem{newwork}
S.~Alexander, T.~Biswas,
``On Exact Non Singular Backgrounds from Heterotic M Theory''
(in preparation)

\bibitem{Biswas2}
 T.~Biswas, R.~Brandenberger, A.~Mazumdar and W.~Siegel,
  ``Non-perturbative gravity, Hagedorn bounce and CMB,''
  arXiv:hep-th/0610274.

\bibitem{NBV}
A.~Nayeri, R.~H.~Brandenberger and C.~Vafa,
  ``Producing a scale-invariant spectrum of perturbations in a Hagedorn phase
  of string cosmology,''
 Phys.\ Rev.\ Lett.\  {\bf 97}, 021302 (2006)   [arXiv:hep-th/0511140].

\bibitem{BNPV2}
R.~H.~Brandenberger, A.~Nayeri, S.~P.~Patil and C.~Vafa,
  ``String gas cosmology and structure formation,''
  arXiv:hep-th/0608121.

\bibitem{RHBrev2}
R.~H.~Brandenberger,
  ``String gas cosmology and structure formation: A brief review,''
  arXiv:hep-th/0702001.


\bibitem{AEGM}
 R.~Allahverdi, K.~Enqvist, J.~Garcia-Bellido and A.~Mazumdar,
  ``Gauge invariant MSSM inflaton,''
  Phys.\ Rev.\ Lett.\  {\bf 97}, 191304 (2006)
  [arXiv:hep-ph/0605035];\\
R.~Allahverdi, A.~Kusenko and A.~Mazumdar,
  ``A-term inflation and the smallness of the neutrino masses,''
  arXiv:hep-ph/0608138. To be published in JCAP.


\bibitem{AEGJM}
R.~Allahverdi, K.~Enqvist, J.~Garcia-Bellido, A.~Jokinen and A.~Mazumdar,
  ``MSSM flat direction inflation: slow roll, stability, fine tunning and
  reheating,''
  JCAP {\bf 0706}, 019 (2007)
  [arXiv:hep-ph/0610134].

\bibitem{AEJM}
R.~Allahverdi, K.~Enqvist, A.~Jokinen and A.~Mazumdar,
  ``Identifying the curvaton within MSSM,''
  JCAP {\bf 0610}, 007 (2006)
  [arXiv:hep-ph/0603255].

\bibitem{MSSM-REV}
 K.~Enqvist and A.~Mazumdar,
  ``Cosmological consequences of MSSM flat directions,''
  Phys.\ Rept.\  {\bf 380}, 99 (2003)
  [arXiv:hep-ph/0209244].

\bibitem{Notari}
A.~Notari and A.~Riotto,
  ``Isocurvature perturbations in the ekpyrotic universe,''
  Nucl.\ Phys.\  B {\bf 644}, 371 (2002)
  [arXiv:hep-th/0205019].

\bibitem{Fabio3}
F.~Finelli,
  ``Assisted contraction,''
  Phys.\ Lett.\  B {\bf 545}, 1 (2002)
  [arXiv:hep-th/0206112].

\bibitem{Turok2}
J.~L.~Lehners, P.~McFadden, N.~Turok and P.~J.~Steinhardt,
  ``Generating ekpyrotic curvature perturbations before the big bang,''
  arXiv:hep-th/0702153.

\bibitem{Ovrut}
E.~I.~Buchbinder, J.~Khoury and B.~A.~Ovrut,
  ``New ekpyrotic cosmology,''
  arXiv:hep-th/0702154;
  arXiv:0706.3903 [hep-th].

\bibitem{Senatore}
P.~Creminelli and L.~Senatore,
  ``A smooth bouncing cosmology with scale invariant spectrum,''
  arXiv:hep-th/0702165.
\bibitem{other}
K.~Koyama, S.~Mizuno and D.~Wands,
  ``Curvature perturbations from ekpyrotic collapse with multiple fields,''
  arXiv:0704.1152 [hep-th];\\
K.~Koyama and D.~Wands,
  ``Ekpyrotic collapse with multiple fields,''
  JCAP {\bf 0704}, 008 (2007)
  [arXiv:hep-th/0703040].

\bibitem{BBM}
S.~Alexander, T.~Biswas, R.~H.~Brandenberger and A.~Mazumdar,
  ``Higher derivative gravity theories, nonsingular bounces and
  ekpyrotic cosmology,''
  arXiv:0707.xxxx.

\bibitem{BK}
R.~H.~Brandenberger and R.~Kahn,
  ``Cosmological Perturbations In Inflationary Universe Models,''
  Phys.\ Rev.\  D {\bf 29}, 2172 (1984).

\bibitem{BST}
J.~M.~Bardeen, P.~J.~Steinhardt and M.~S.~Turner,
  ``Spontaneous Creation Of Almost Scale - Free Density Perturbations In An
  Inflationary Universe,''
  Phys.\ Rev.\  D {\bf 28}, 679 (1983).

\bibitem{Sasaki}
M.~Sasaki,
  ``Large Scale Quantum Fluctuations in the Inflationary Universe,''
  Prog.\ Theor.\ Phys.\  {\bf 76}, 1036 (1986).

\bibitem{Mukhanov}
V.~F.~Mukhanov,
  ``Quantum Theory of Gauge Invariant Cosmological Perturbations,''
  Sov.\ Phys.\ JETP {\bf 67}, 1297 (1988)
  [Zh.\ Eksp.\ Teor.\ Fiz.\  {\bf 94N7}, 1 (1988)].

\bibitem{Lyth}
D.~H.~Lyth,
  ``Large Scale Energy Density Perturbations And Inflation,''
  Phys.\ Rev.\  D {\bf 31}, 1792 (1985).

\bibitem{Pinho}
  E.~J.~C.~Pinho and N.~Pinto-Neto,
  Phys.\ Rev.\  D {\bf 76}, 023506 (2007)
  [arXiv:hep-th/0610192].

\end{thebibliography}
\end{document}